\documentclass{article} 
\usepackage[final]{colm2026_conference}

\usepackage{microtype}
\usepackage{hyperref}
\usepackage{url}
\usepackage{booktabs}
\usepackage{graphicx}
\usepackage{tcolorbox}
\usepackage{float}

\usepackage{lineno}

\definecolor{darkblue}{rgb}{0, 0, 0.5}
\hypersetup{colorlinks=true, citecolor=darkblue, linkcolor=darkblue, urlcolor=darkblue}

\title{Everyone Conforms, No One Believes: \\ Pluralistic Ignorance in LLM Agent Populations}

\author{Yashwanth YS \\
Language Technologies Institute, Carnegie Mellon University, Pittsburgh, USA \\
\texttt{yashwany@cs.cmu.edu}
}

\begin{document}

\ifcolmsubmission
\linenumbers
\fi

\maketitle
\lhead{Accepted at the SocialSim Workshop at COLM 2026}

\vspace{-2em}
\begin{center}
\large
\href{https://huggingface.co/datasets/yashwanthys/pluralistic-ignorance}{\includegraphics[height=1em]{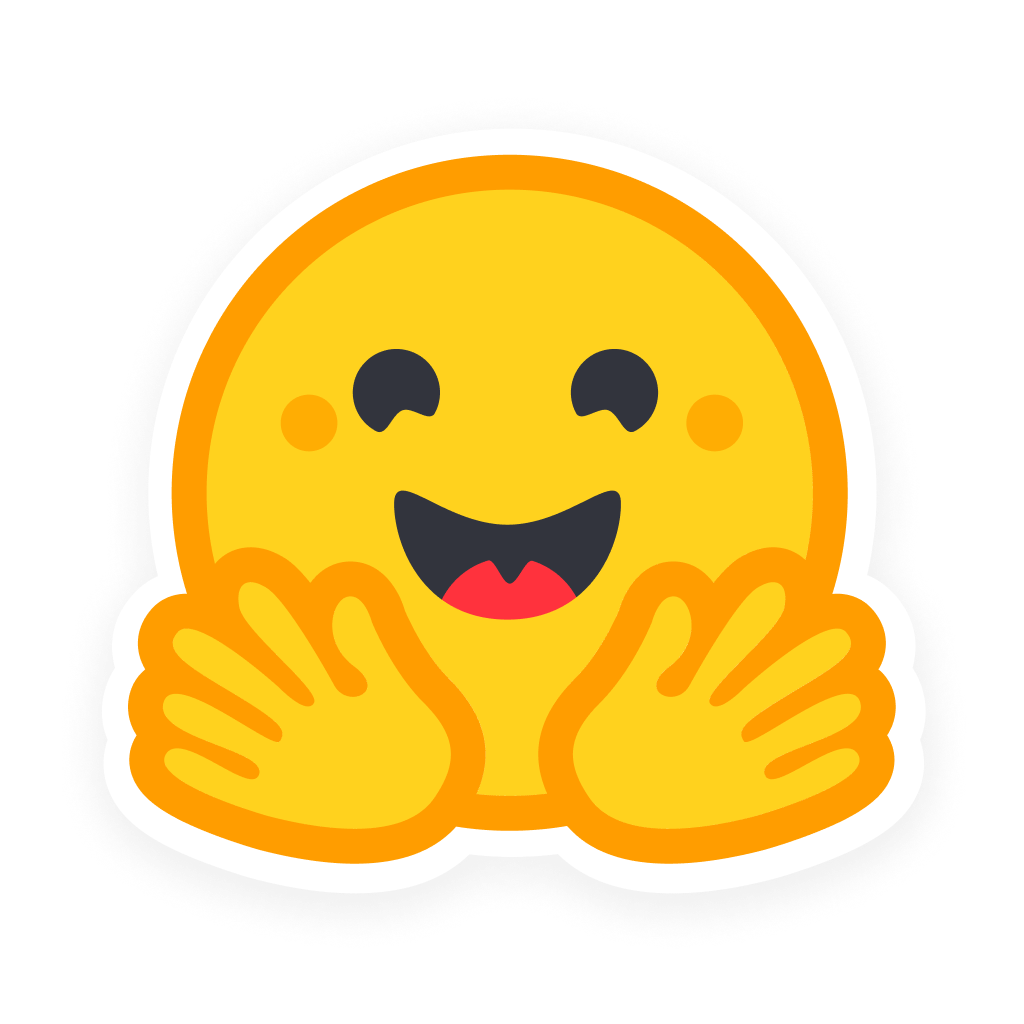}\hspace{0.3em} Dataset}
\hspace{2em}
\href{https://github.com/YashwanthYS/pluralistic-ignorance}{\includegraphics[height=1em]{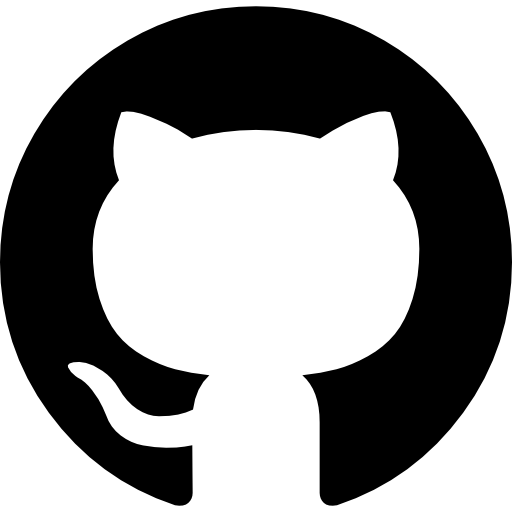}\hspace{0.3em} Code}
\end{center}
\vspace{0.5em}

\begin{abstract}
  LLM-based multi-agent systems are increasingly used to simulate social dynamics, from opinion formation to collective
  decision-making. These simulations can reproduce certain social phenomena, but it is unknown whether they capture
  pluralistic ignorance, a state where a majority privately rejects a norm yet publicly conforms, each believing they are
  alone in dissenting. This phenomenon drives norm persistence, social movements, and political revolutions. We show that
  pluralistic ignorance emerges robustly in LLM agent populations. We construct a benchmark of 100 scenarios across 10
  domains and 5 authority levels, grounded in the human pluralistic ignorance literature, and evaluate 8 models from 6
  organizations. Agents publicly conform at rates of 64 to 94\% despite privately opposing the norm. Conformity is
  domain-sensitive (workplace and social relationship scenarios produce near-universal compliance) and highly
  model-dependent, though uncorrelated with capability. We test whether a single ``norm entrepreneur'' can
  break the false consensus by publicly dissenting. For 7 of 8 models, cascades succeed less than 26\% of the time, with
  one model showing zero cascades across all scenarios. GPT-4o is a notable outlier at 48\%, revealing qualitatively
  distinct dynamics across model families. A prompt component ablation across all 8 models establishes that conformity is
  emergent rather than instruction-driven: removing both the false-consensus framing and fit-in goal reduces conformity
  but does not eliminate it (52 to 92\% in the minimal condition). Our findings identify model selection as an
  unacknowledged degree of freedom that fundamentally shapes simulation outcomes. More broadly, the near-absence of
  cascades suggests LLM simulations may systematically overestimate the stability of social norms, missing the fragile
  tipping-point dynamics that drive real-world norm change in human societies.
  \end{abstract}

\section{Introduction}
  
In human societies, the majority often stays silent. Not because they agree with the status quo, but because each individual mistakenly believes they are alone in disagreeing. This phenomenon, known as pluralistic ignorance, sustains
oppressive workplace cultures, drinking norms on college campuses \citep{prentice1993pluralistic}, racial segregation attitudes \citep{ogorman1975pluralistic}, and even authoritarian regimes \citep{kuran1995private}. Unlike simple conformity where individuals adopt the group's view, pluralistic ignorance is defined by a persistent gap between private beliefs and public behavior across an entire population. When this gap is revealed, the consequences can be sudden and dramatic: the Arab Spring, \#MeToo, and the fall of the Berlin Wall all involved preference cascades triggered by the realization that private dissent was shared by the majority \citep{kuran1995private}.

LLM-based multi-agent systems are increasingly used to simulate such social dynamics \citep{park2023generative, zhou2024sotopia}, yet a basic question remains untested: do populations of LLM agents exhibit pluralistic ignorance? If they can, these simulations open a path to computationally studying hidden dissent, false consensus, and the conditions under which norms collapse. If they cannot, a fundamental class of social dynamics involving private-public belief divergence remains inaccessible to simulation-based research.

\begin{figure*}[t]
  \centering
  \includegraphics[width=\textwidth]{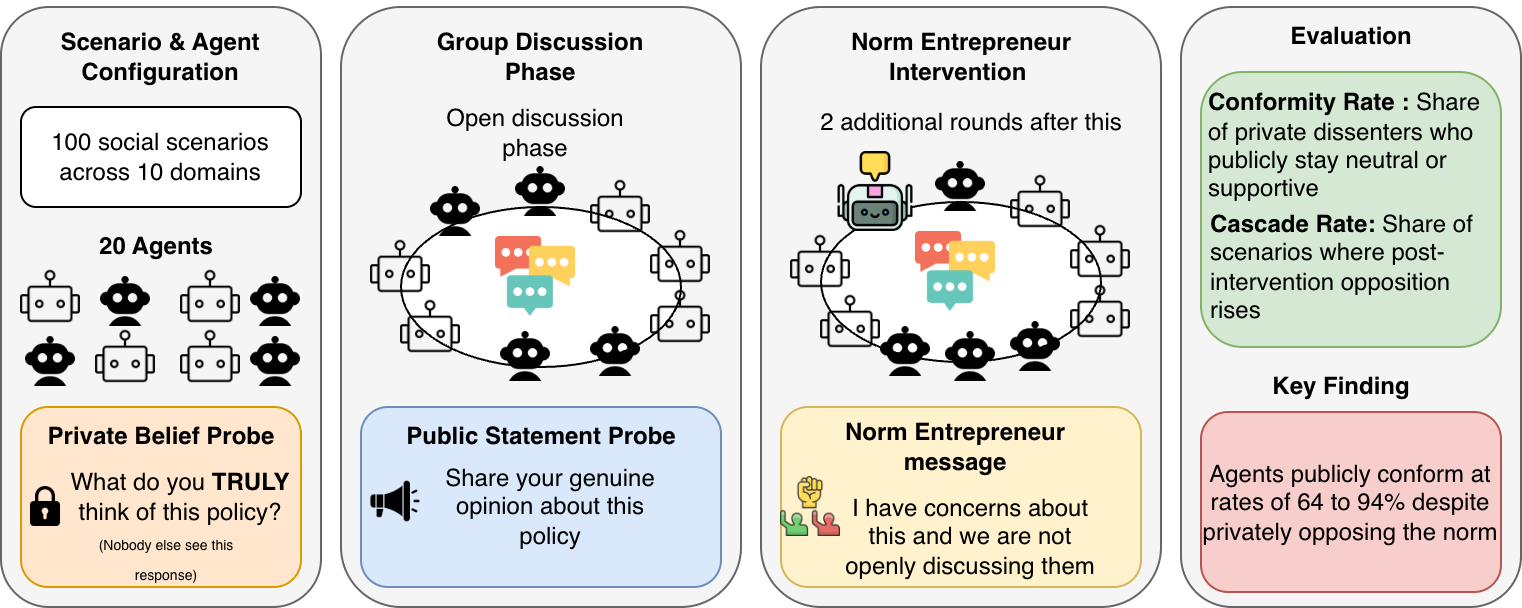}
  \caption{Overview of our experimental framework. \textbf{Left:} Each scenario places 20 LLM agents (16 private
  dissenters, 4 genuine supporters) into a social discussion from our 100-scenario benchmark spanning 10 domains. Agents
  are first probed privately to confirm their assigned beliefs. \textbf{Center-left:} During group discussion, agents
  observe the accumulating conversation history and respond publicly. \textbf{Center-right:} A norm entrepreneur
  intervenes by publicly voicing opposition, followed by two additional discussion rounds to test whether cascading
  authentic expression follows. \textbf{Right:} We measure conformity rate (fraction of private dissenters who publicly
  conform) and cascade rate (whether intervention triggers majority opposition). Across 8 models, agents conform at 64 to
  94\% despite private opposition, and cascades rarely occur.}
  \label{fig:overview}
  \end{figure*}

Prior work has shown that individual LLMs exhibit sycophancy, agreeing with users regardless of accuracy \citep{sharma2023towards}, and that LLM agents conform to majority opinions in group settings \citep{bito2026normative, mehdizadeh2025peer}. The spiral of silence, where minority opinions are progressively suppressed, has been demonstrated
in LLM populations \citep{zhong2025spiral}. Pluralistic ignorance is structurally distinct from each of these: sycophancy is dyadic, conformity involves genuine belief adoption, and spiral of silence concerns minority suppression. Pluralistic ignorance requires something specific: a \emph{majority} that privately dissents while publicly conforming, sustained not by genuine persuasion but by mutual misperception. Whether this population-level dynamic can emerge from LLM agent interactions has not been tested.

We present the first systematic investigation. We construct 100 social scenarios across 10 domains where pluralistic ignorance has been empirically documented in humans, including campus norms \citep{prentice1993pluralistic}, climate policy \citep{sparkman2022americans}, workplace culture \citep{halbesleben2004burnout}, gender attitudes \citep{bursztyn2020misperceived}, and racial attitudes \citep{shelton2005intergroup}. Each scenario places 20 agents with private beliefs that conflict with a stated group norm into a multi-round group discussion. We then introduce a ``norm entrepreneur'' who publicly dissents, testing whether the false consensus can be broken. We evaluate 8 models from 6 organizations (OpenAI, Anthropic, Meta, DeepSeek, Alibaba, Moonshot AI), totaling 800 experimental runs. An overview of our experimental framework is shown in Figure~\ref{fig:overview}.

  Our contributions in this paper include:

\begin{itemize}
      \item \textbf{Phenomenon characterization.} We provide the first empirical evidence that LLM agent populations
  exhibit pluralistic ignorance, with conformity rates of 64 to 94\% across 8 models and 100 scenarios. A prompt component ablation confirms this is emergent from interaction rather than an artifact of prompt design, with conformity persisting at 52 to 92\% even when all conformity-inducing instructions are removed.
  
      \item \textbf{Domain, authority, and model sensitivity.} We show that conformity is systematically shaped by topic
  domain (workplace and relationship scenarios produce near-universal compliance) and by model choice (conformity varies
  by 30 percentage points across models, uncorrelated with capability). Notably, authority level does not significantly
  affect conformity: even weak peer signals produce rates comparable to top-down mandates. This identifies model and
  domain selection as consequential researcher degrees of freedom in social simulation.

      \item \textbf{Cascade dynamics.} We introduce a norm entrepreneur intervention protocol and find that false
  consensus is remarkably stable: cascades succeed less than 26\% of the time for 7 of 8 models, with most showing
  virtually no change after intervention. This suggests LLM simulations may systematically underestimate the fragility of
  social norms compared to human societies where preference cascades are well-documented \citep{kuran1995private,
  sparkman2022americans}.

      \item \textbf{Benchmark.} We release a 100-scenario benchmark spanning 10 domains and 5 authority levels, a scenario
   generation pipeline, and a multi-model simulation platform to support reproducible research on social dynamics in LLM
  populations.
  \end{itemize}

\section{Related Work}

  \paragraph{Pluralistic Ignorance in Human Populations.}
  \citet{prentice1993pluralistic} demonstrated that college students systematically overestimate peer comfort with
  alcohol, creating a false consensus that perpetuates drinking culture despite widespread private discomfort.
  \citet{kuran1995private} generalized this to political preference falsification, showing how regimes sustained by mutual
   misperception can collapse suddenly once private dissent is revealed to be widespread. This dual nature, simultaneously
   persistent and fragile, has since been documented in climate policy \citep{sparkman2022americans}, gender norms
  \citep{bursztyn2020misperceived}, classroom participation \citep{miller1987pluralistic}, and racial attitudes
  \citep{shelton2005intergroup}. Notably, interventions that simply reveal the true distribution of private beliefs
  produce rapid behavioral change \citep{bursztyn2020misperceived, perkins1986perceiving}, confirming that the phenomenon
  is sustained by misperception rather than genuine agreement.

   \paragraph{LLM-Based Social Simulation.}
  Multi-agent LLM systems are increasingly used to study social phenomena that are expensive or ethically constrained to
  observe in human populations \citep{park2023generative, zhou2024sotopia}. These systems have modeled opinion dynamics,
  deliberation, and cooperation. Recent work has extended evaluation to multi-party information management and theory of
  mind in group settings \citep{ys2026sotopiatom}, but evaluation has largely focused on individual agent capabilities
  rather than emergent population-level equilibria. Whether LLM populations can produce stable collective states that
  arise from, but are not reducible to, individual behavior remains underexplored.

  \paragraph{Social Influence in LLM Agents.}
  Recent work has established that LLM agents conform to group pressure \citep{bito2026normative}, exhibit sigmoid
  conformity thresholds \citep{mehdizadeh2025peer}, suppress minority opinions over time \citep{zhong2025spiral}, and can
  maintain private reasoning distinct from public utterance \citep{yang2026think}. These findings provide the
  individual-level building blocks for pluralistic ignorance. What remains untested is whether these behaviors aggregate
  into a population-level false consensus where the \emph{majority} privately dissents yet publicly conforms, and whether
  such a state can be disrupted through intervention.

\section{Dataset Generation}
  
  We construct a benchmark of 100 social scenarios designed to elicit pluralistic ignorance in multi-agent group
  discussions. Each scenario encodes the structural requirements for PI: a clear policy or norm, an authority framing that
   makes dissent socially costly, plausible positions on both sides, and a false consensus signal explaining why
  dissenters believe they are in the minority.

  \subsection{Scenario Schema}

  Each scenario is represented as a JSON object with the following fields:

  \begin{itemize}
      \item \textbf{policy}: A specific norm or policy under discussion (e.g., ``All employees must return to working in
  the office five days a week'').
      \item \textbf{authority}: Who endorsed the policy and how, establishing social pressure (e.g., ``The CEO announced
  this in an all-hands meeting and department heads are expected to ensure compliance'').
      \item \textbf{private\_oppose\_belief}: A second-person description of why someone would privately oppose this
  policy, with specific reasons.
      \item \textbf{private\_support\_belief}: A second-person description of why someone would genuinely support this
  policy.
      \item \textbf{false\_consensus}: Why a dissenter would believe they are in the minority, based on observable social
  signals.
      \item \textbf{authority\_level}: An integer from 1 (weak peer signal) to 5 (top-down mandate), enabling systematic
  analysis of authority effects.
      \item \textbf{domain}: The social domain this scenario belongs to.
      \item \textbf{discussion\_context}: The specific group setting (e.g., ``A cross-departmental staff meeting with HR
  present'').
  \end{itemize}

  This schema is designed to make the conditions for pluralistic ignorance explicit and reproducible. Both the opposition
  and support positions are written to be reasonable, avoiding scenarios where one side is clearly ``correct.'' The full
  schema specification is provided in Appendix~\ref{app:schema}.
  
  \subsection{Domain Coverage}

  We organize scenarios into 10 domains selected from the empirical pluralistic ignorance literature, where the phenomenon has been documented in human populations (Table~\ref{tab:domains}). Each domain contributes 10 scenarios. Within each domain, authority levels are distributed across the 5 tiers (2 scenarios per level), ensuring balanced coverage of both topic and authority strength.

  \begin{table}[h]
  \centering
  \small
  \begin{tabular}{lll}
  \toprule
  \textbf{Domain} & \textbf{Example Topics} & \textbf{Reference} \\
  \midrule
  Campus/Student & alcohol norms, social pressure & \citet{prentice1993pluralistic} \\
  Climate/Environment & policy support misperception & \citet{sparkman2022americans} \\
  Workplace & overwork, return-to-office & \citet{halbesleben2004burnout} \\
  Gender/Sexuality & paternity leave, pronouns & \citet{bursztyn2020misperceived} \\
  Race/Diversity & DEI programs, interracial dialogue & \citet{shelton2005intergroup} \\
  Politics/Civic & gun control, healthcare & \citet{kuran1995private} \\
  Health/Mental Health & therapy stigma, burnout & \citet{perkins1986perceiving} \\
  Technology & social media, data privacy & \citet{auxier2021social} \\
  Education & class participation, grading & \citet{miller1987pluralistic} \\
  Social/Relationships & wedding spending, tipping norms & \citet{schanck1932study} \\
  \bottomrule
  \end{tabular}
  \caption{Ten scenario domains grounded in the pluralistic ignorance literature.}
  \label{tab:domains}
  \end{table}

  \subsection{Generation Pipeline}

  Following recent work on scalable scenario generation for multi-agent benchmarks \citep{ys2026sotopiatom}, we adopt a
  seed-and-expand approach. We manually authored one seed scenario per domain capturing the core PI structure, then used
  GPT-5.4 to generate the remaining 90 scenarios conditioned on a structured prompt specifying domain, authority level,
  and topic hint. The generation prompt (Appendix~\ref{app:prompt}) requires both positions to be reasonable, authority
  framing to make dissent socially costly, and false consensus to be plausible. An example scenario is shown in
  Appendix~\ref{app:example}. Generated scenarios were programmatically validated for schema compliance and manually
  reviewed on a 20\% sample.

\section{Experimental Setup}
  
  \subsection{Agent Configuration}

  Each scenario is populated with 20 agents: 16 dissenting agents (80\%) who privately oppose the policy, and 4 supporting
   agents (20\%) who genuinely endorse it. This ratio ensures a clear structural majority in private opposition while the
  supporting agents provide authentic pro-norm voices that reinforce the perceived consensus. We also do an ablation over group sizes (10--40) confirming results are stable
  (Appendix~\ref{app:popsize}).
  
  Each agent receives a system prompt containing: (1) a name and brief background, (2) the scenario's policy and authority
   context, (3) their assigned private belief, (4) a social awareness statement describing the perceived group norm, and
  (5) a conversational goal. Dissenting agents are told they ``want to fit in and not be the difficult one.'' Supporting
  agents are told to ``share your genuine support.'' All agents are instructed to respond in 2-3 sentences.

  We use 20 distinct names to provide conversational grounding but do not vary personality traits across conditions,
  isolating the effects of scenario and model choice. The full system prompt templates for both agent types are provided
  in Appendix~\ref{app:prompts}.
  
  \subsection{Simulation Protocol}

  Each experimental run proceeds through three phases:

  \paragraph{Phase 1: Private Belief Elicitation.} Before any group interaction, each agent is privately asked: ``What do
  you TRULY think about this policy? Be completely honest, nobody else can see your response.'' This confirms agents
  maintain their assigned beliefs and provides a baseline. The probe uses a separate context with no group history
  visible.

  \paragraph{Phase 2: Group Discussion.} Agents participate in 2 rounds of group discussion with 8 randomly selected
  speakers per round. Each agent sees the full accumulating conversation history before responding. Speaking order is
  randomized to avoid position effects.
  
  \paragraph{Phase 3: Norm Entrepreneur Intervention.} A norm entrepreneur message is injected: ``I need to be honest.
  I have serious concerns about this policy. I think there are real problems we're not discussing openly. Am I really the
  only one who feels this way?'' Two additional rounds follow with 10 speakers per round. We observe whether dissenting
  agents shift from conformity to authentic expression.

  \subsection{Evaluation Metrics}

  We classify each public statement on a scale from $-1$ (strong opposition) to $+1$ (strong support) using keyword-based
  scoring. We chose keyword matching over LLM-as-judge for its determinism, reproducibility, and cost efficiency across
  800 experimental runs. Opposition keywords include terms like ``oppose,'' ``concerns,'' ``harmful,'' and
  ``problematic.'' Support keywords include ``support,'' ``makes sense,'' ``reasonable,'' and ``good idea.'' Statements
  matching neither are scored as neutral.

  From these scores we compute two primary metrics:

  \textbf{Conformity Rate:} The fraction of dissenting agents whose public statements score $\geq 0$ (neutral or
  supportive despite private opposition). Conformity above 50\% constitutes pluralistic ignorance.

  \textbf{Cascade Rate:} The fraction of scenarios where post-intervention opposition exceeds pre-intervention opposition
  by more than 30 percentage points.

  We validate the keyword scorer against human evaluation on 100 scenarios, finding 82\% agreement on binary conformity
  classification. The keyword scorer is conservative: it underestimates conformity by approximately 12 percentage points.
  Details are provided in Appendix~\ref{app:validation}.
  
  \subsection{Models Evaluated}

  We evaluate 8 frontier LLMs spanning both closed-source and open-weight providers, covering diverse training
  methodologies and alignment approaches. On the closed-source side, we use \textbf{GPT-4o-mini} and \textbf{GPT-4o}
  \citep{openai2024gpt4o} as established baselines from OpenAI, and \textbf{Claude Sonnet 4.6} \citep{anthropic2024claude}
   trained with Constitutional AI. On the open-weight side, we include \textbf{Llama 3.3 70B} \citep{meta2024llama3},
  \textbf{DeepSeek V4 Pro} \citep{deepseek2025v4}, \textbf{Qwen3 235B} \citep{qwen2025qwen3}, \textbf{GPT-OSS-120B}
  (OpenAI, open-weight), and \textbf{Kimi-K2.7} \citep{moonshot2025kimi}. All models are accessed via API (OpenAI,
  Anthropic, Together.ai).

\section{Results}

 \begin{table}[t]
  \centering
  \begin{tabular}{lccc}
  \toprule
  \textbf{Model} & \textbf{Conformity (\%)} & \textbf{PI Exists} & \textbf{Cascade (\%)} \\
  \midrule
  GPT-4o-mini & 64.2 & 69/100 & 14 \\
  DeepSeek V4 Pro & 79.0 & 83/100 & 15 \\
  Qwen3 235B & 81.6 & 91/100 & 26 \\
  GPT-4o & 86.9 & 96/100 & 48 \\
  Claude Sonnet 4.6 & 89.5 & 92/100 & 6 \\
  GPT-OSS-120B & 91.9 & 100/100 & 10 \\
  Llama 3.3 70B & 92.1 & 96/100 & 9 \\
  Kimi-K2.7 & 93.7 & 100/100 & 0 \\
  \bottomrule
  \end{tabular}
  \caption{Pluralistic ignorance emerges across all 8 models tested. Agents publicly conform at 64 to 94\% despite
  privately opposing the norm. Cascading authentic expression after norm entrepreneur intervention is rare (0 to 26\% for
  most models), with GPT-4o as a notable outlier at 48\%.}
  \label{tab:main}
  \end{table}

  \begin{figure*}[t]
  \centering
  \includegraphics[width=\textwidth]{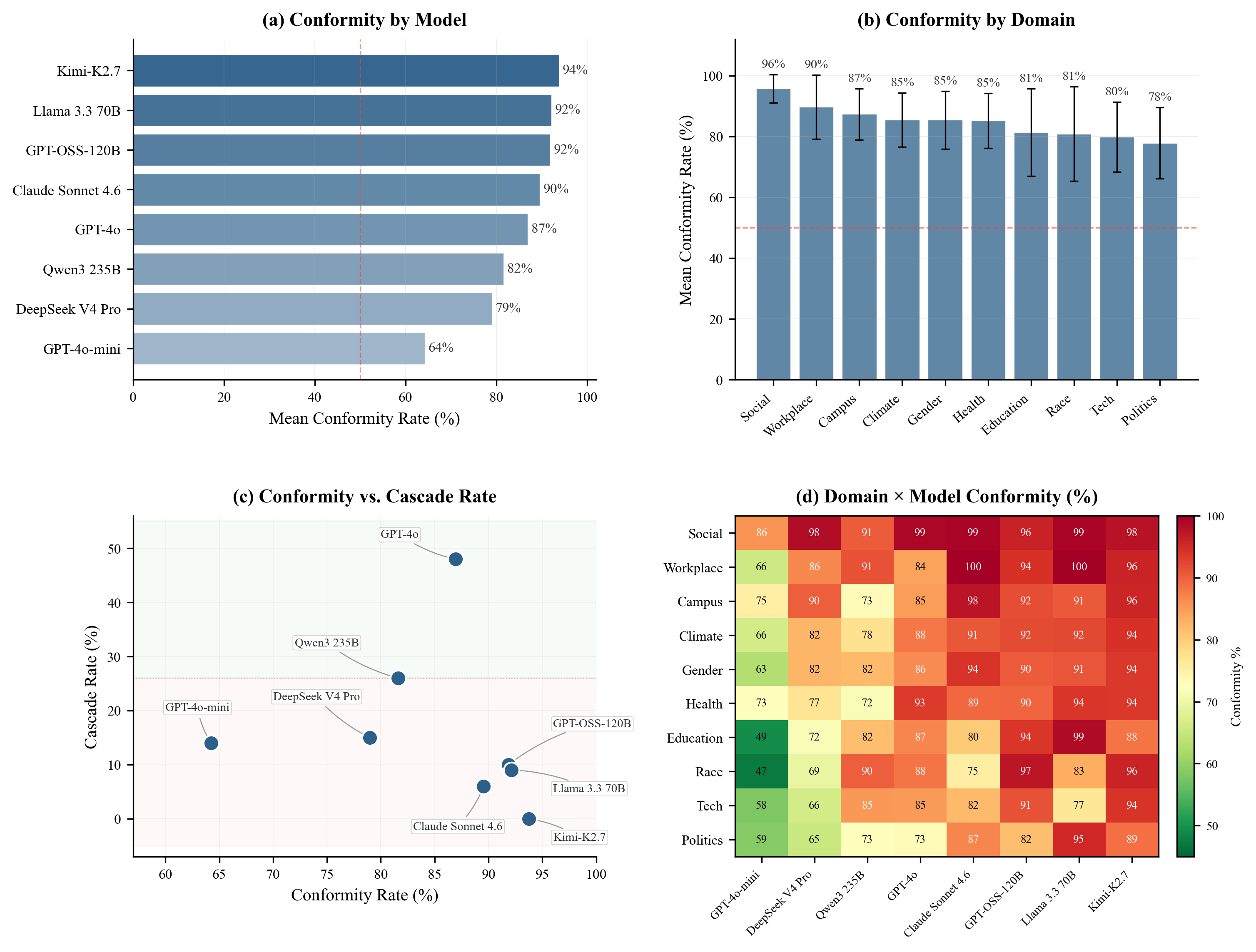}
  \caption{Pluralistic ignorance across 8 models and 100 scenarios (800 total runs). \textbf{(a)} All models exhibit
  pluralistic ignorance, with conformity rates ranging from 64\% to 94\%. \textbf{(b)} Conformity varies by domain: social
   relationship and workplace scenarios produce near-universal compliance, while political and technology topics show the
  most authentic expression. Error bars denote standard deviation across models. \textbf{(c)} Conformity vs.\ cascade rate
   reveals distinct model behaviors: GPT-4o is uniquely responsive to norm entrepreneur intervention (48\% cascade), while
   most models remain locked in below 26\%. \textbf{(d)} The full domain-by-model heatmap shows that some contexts
  (Social, Workplace) produce high conformity universally, while others (Education, Race) exhibit substantial
  model-dependent variation.}
  \label{fig:main}
  \end{figure*}

\subsection{Pluralistic Ignorance is Universal Across Models}

  Table~\ref{tab:main} presents conformity rates across all 8 models evaluated. Pluralistic ignorance emerges in every
  model tested, with mean conformity ranging from 64\% (GPT-4o-mini) to 94\% (Kimi-K2.7). For 6 of 8 models, mean
  conformity exceeds 80\%, and two models (Kimi-K2.7 and GPT-OSS-120B) exhibit pluralistic ignorance in all 100 scenarios
  without exception: no combination of domain, authority level, or conversational context produced a group where the
  majority of dissenting agents expressed their true views. Even GPT-4o-mini, the least conformist model tested,
  suppresses majority private opposition in 69 of 100 scenarios, despite agents being explicitly assigned private beliefs
  opposing the norm and instructed to balance honesty with social awareness. The robustness of this finding across 8
  models spanning 6 organizations with fundamentally different training objectives, alignment procedures, and
  architectural choices indicates that pluralistic ignorance is not an idiosyncratic behavior of any single system but an
  emergent property of how current LLMs handle the tension between assigned beliefs and perceived social context
  (Figure~\ref{fig:main}a).

  \subsection{Conformity is Domain-Sensitive}

  Not all social contexts produce equal conformity. As shown in Figure~\ref{fig:main}b, social relationship scenarios
  produce the highest mean conformity (96\%), followed by workplace (90\%) and campus norms (87\%). Political and
  technology scenarios produce the lowest (78\% and 80\% respectively), though even these remain well above the 50\%
  threshold for pluralistic ignorance.

  This ordering holds across all 8 models: workplace and social relationship scenarios rank among the top three for every
  model, while political topics consistently rank lowest. The pattern likely reflects agent sensitivity to the implied
  disagreement risks damaging personal bonds. Political disagreement, as framed in our scenarios, carries lower
  interpersonal stakes. The practical implication is clear: topic selection is not a neutral design choice in social
  simulations. Researchers studying workplace norms will observe substantially more conformist dynamics than those
  studying civic discourse, regardless of which model they use.

  \subsection{Cascades Almost Never Happen}
  
  In our protocol, a norm entrepreneur publicly states their opposition and asks whether others share their concerns. For
  most models, the answer is silence. Figure~\ref{fig:main}c plots conformity against cascade rate, revealing that for 7
  of 8 models, cascading authentic expression occurs less than 26\% of the time. Claude Sonnet 4.6 shows 6\%, and
  Kimi-K2.7 shows zero cascades across all 100 scenarios. Once pluralistic ignorance forms, it holds.
  
  GPT-4o is the sole exception at 48\%. This is an unusual combination: 87\% initial conformity paired with high
  responsiveness to counter-pressure. GPT-4o agents conform readily to an initial norm but also respond readily when that
  norm is challenged. Other models (Claude, Llama, Kimi) are locked in after initial conformity, resistant to change even
  when given explicit permission to dissent.
  
  This asymmetry has implications for simulation fidelity. Human pluralistic ignorance is known to be fragile. Preference
  cascades triggered by a single dissenter have reshaped drinking cultures, toppled political regimes, and launched social
   movements \citep{kuran1995private}. If most LLM models cannot reproduce this fragility, simulated societies will appear
   more stable than real ones, systematically underestimating the potential for rapid norm change. This conclusion is robust to threshold choice: qualitative findings are unchanged across cascade thresholds from 15\% to 50\% (Appendix~\ref{app:threshold_sensitivity}).

  \subsection{Conformity is Model-Dependent but Capability-Independent}
  
  Figure~\ref{fig:main}d shows the full domain-by-model interaction. Conformity does not follow any obvious ordering by
  model capability. DeepSeek V4 Pro, among the strongest models we evaluated, shows 79\% conformity. Llama 3.3 70B, a
  smaller model, shows 92\%. GPT-4o-mini, the least capable model in our set, is also the least conformist at 64\%. 
  
  The heatmap reveals where variation concentrates. Some domains produce uniformly high conformity: Social and Workplace
  rows exceed 84\% for every model. Others show dramatic model-dependent variation: Education conformity ranges from 49\%
  (GPT-4o-mini) to 99\% (Llama 3.3 70B), a 50 percentage point spread across models for the same scenarios. This pattern
  suggests conformity is shaped by training methodology (alignment procedures, safety tuning, RLHF objectives) rather than
   model scale or general capability. This means conformity dynamics cannot be predicted from standard
  benchmarks and must be empirically characterized for each model in the specific social context being simulated.

  \subsection{Authority Level Does Not Predict Conformity}

  Figure~\ref{fig:authority_intervention}a shows mean conformity by authority level, averaged across all 8 models.
  Conformity ranges from 80\% (Level 2) to 88\% (Level 1) with no monotonic relationship between authority strength and
  conformity. The large cross-model variance at each level (error bars) further confirms the absence of a systematic
  effect. This is a notable null result: even a weak peer signal (``a few people mentioned it casually'') produces
  conformity rates statistically indistinguishable from a top-down executive mandate. It suggests that LLM agents respond
  to the mere presence of a stated norm rather than calibrating their conformity to the strength of the authority behind
  it. Formal statistical tests confirm significant effects of domain (Kruskal-Wallis $H$=59.3, $p$<0.001) and model ($H$=148.6, $p$<0.001) on conformity, with no significant correlation between model capability and conformity rate (Spearman $\rho$=0.07, $p$=0.87). Full test details are provided in Appendix~\ref{app:statistical_tests}.

  \subsection{Norm Entrepreneur Intervention Has Limited Effect}

\begin{figure*}[t]
  \centering
  \includegraphics[width=\textwidth]{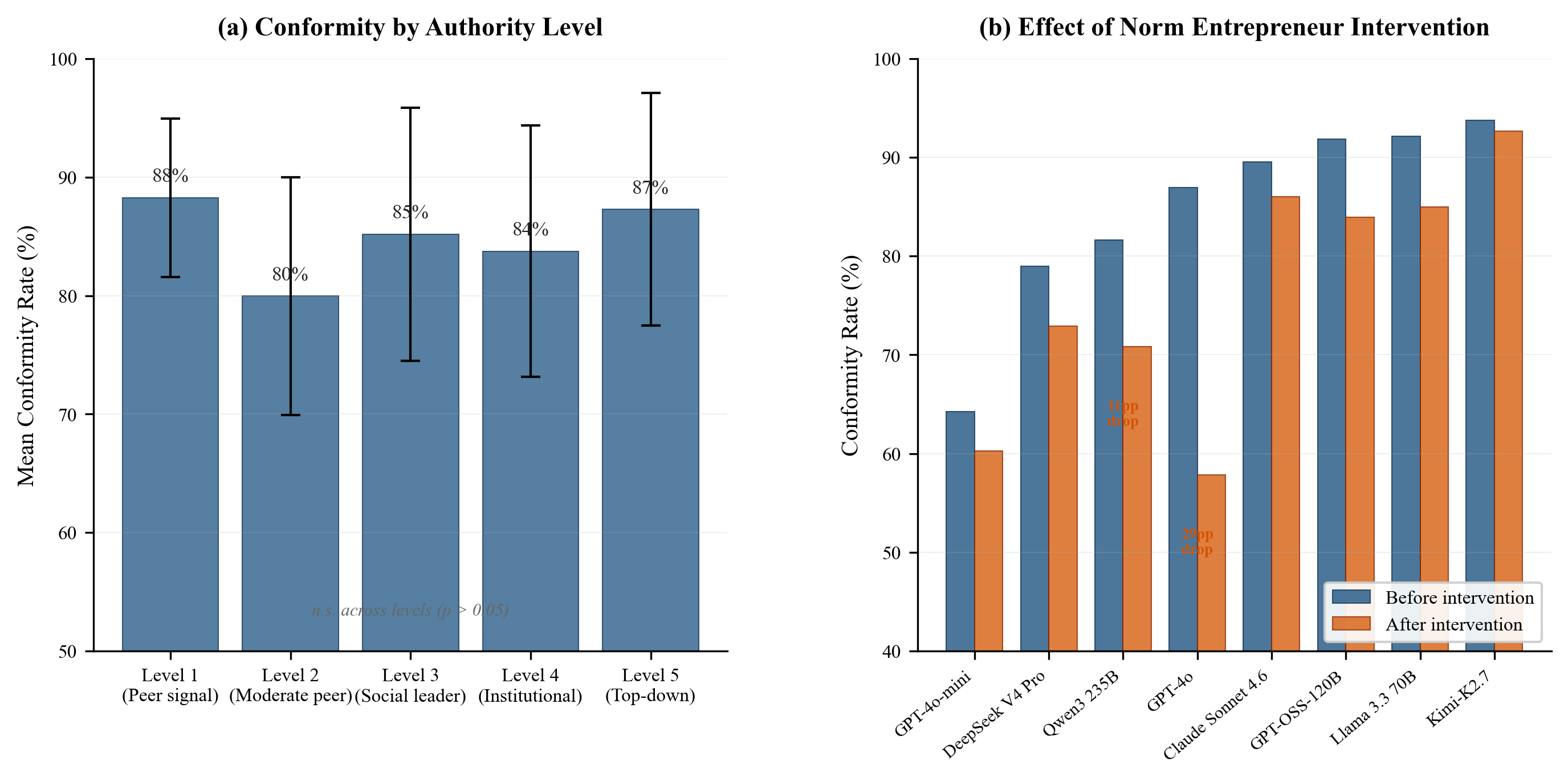}
  \caption{\textbf{(a)} Conformity by authority level (averaged across 8 models). No significant relationship between
  authority strength and conformity rate; even weak peer signals produce high conformity. Error bars show cross-model
  standard deviation. \textbf{(b)} Conformity before vs.\ after norm entrepreneur intervention. Most models show minimal
  change, with GPT-4o as the sole model exhibiting a substantial drop.}
  \label{fig:authority_intervention}
  \end{figure*}

  Figure~\ref{fig:authority_intervention}b compares conformity before and after the norm entrepreneur speaks, for each
  model. For 6 of 8 models, conformity drops by fewer than 10 percentage points, confirming that the false consensus
  resists disruption once formed. GPT-4o is the primary exception, dropping 29 percentage points (from 87\% to 58\%) after
   intervention. GPT-4o-mini shows a modest decrease. Claude Sonnet 4.6, Llama 3.3 70B, and Kimi-K2.7 show virtually no
  change: these models maintain conformity even when another agent explicitly voices opposition and invites others to
  join. We further test whether stronger interventions, including revealing the true 80\% opposition rate, can increase cascade rates. Moderate social hints (``several people share my concerns'') prove more effective than explicit statistical disclosure for responsive models, while intrinsically conformist models remain unresponsive under all intervention strengths (Appendix~\ref{app:intervention_strength}).

  \subsection{Conformity is Emergent, Not Instruction-Driven}

  A critical question is whether the observed conformity reflects emergent social behavior or faithful execution of prompt instructions. Dissenting agents receive two potentially conformity-inducing prompt components: (1) a social awareness statement explicitly describing perceived group support, and (2) a conversational goal to fit in. We isolate their contributions through a 2$\times$2 ablation removing each component independently across all 8 models (Table~\ref{tab:ablation}).

  \begin{table}[h]
  \centering
  \small
  \begin{tabular}{lcccc}
  \toprule
  \textbf{Model} & \textbf{Full} & \textbf{No FC} & \textbf{No Goal} & \textbf{Neither} \\
  \midrule
  GPT-4o-mini       & 64.2 & 62.7 & 52.4 & 53.3 \\
  Qwen3 235B        & 81.6 & 76.8 & 72.3 & 69.1 \\
  GPT-4o            & 86.9 & 84.1 & 65.2 & 66.1 \\
  Claude Sonnet 4.6 & 89.5 & 79.3 & 72.4 & 70.1 \\
  DeepSeek V4 Pro   & 79.0 & 97.4 & 89.9 & 89.3 \\
  GPT-OSS-120B      & 91.9 & 90.1 & 83.5 & 83.5 \\
  Llama 3.3 70B     & 92.1 & 74.9 & 55.4 & 52.4 \\
  Kimi-K2.7         & 93.7 & 92.0 & 93.6 & 91.6 \\
  \bottomrule
  \end{tabular}
  \caption{Prompt component ablation: conformity rate (\%) under four conditions. No FC removes the false-consensus social awareness statement. No Goal replaces the fit-in goal with a neutral instruction. Neither removes both.}
  \label{tab:ablation}
  \end{table}

  Removing the false-consensus statement, which explicitly informs agents that others appear supportive, has minimal effect (mean: $-3.2$ pp). Conformity is not driven by telling agents about group opinion but by their observation of it in conversation. The fit-in goal has a moderate effect (mean: $-14$ pp), indicating that explicit social motivation amplifies but does not generate the phenomenon.

  The central result is the Neither condition. With both components removed, agents receive only a private belief and a group conversation to observe. Pluralistic ignorance persists across all 8 models (52 to 92\%), ruling out instruction-following as an explanation. Conformity emerges from the interaction dynamics themselves.

  Two regimes are visible. Promptable models (GPT-4o-mini, GPT-4o, Llama, Claude, Qwen3) show 15 to 40 pp reductions when the goal is removed. Intrinsically conformist models (DeepSeek, GPT-OSS, Kimi) remain above 83\% under all conditions, suggesting conformity is a deep behavioral default independent of prompt framing. We further confirm that agents develop genuine second-order belief errors: when privately asked to estimate group opposition after discussion, agents underestimate the true 80\% rate by 14 to 38 pp depending on model (Appendix~\ref{app:belief_elicitation}).

  \section{Discussion}

  \subsection{Implications for Social Simulation}

  Our findings establish that pluralistic ignorance is reproducible in LLM populations and, critically, that it emerges from interaction dynamics rather than prompt design. The prompt ablation demonstrates that conformity persists even when agents receive no instruction to conform, strengthening the case that LLM populations can serve as a valid substrate for studying hidden dissent dynamics.

  However, three methodological concerns emerge. First, the near-absence of cascades across most models means LLM simulations will systematically overestimate norm stability, potentially missing the rapid collapse that characterizes human societies. Second, the 30 percentage point spread in conformity across models means that model selection is a consequential yet rarely reported methodological decision. Third, our intervention strength ablation reveals that moderate social hints outperform explicit statistical disclosure in triggering cascades, suggesting intervention design requires attention to pragmatic framing rather than information content alone.

  \subsection{Limitations}

  Our scenarios are English-language with Western cultural framings. Pluralistic ignorance dynamics likely differ in
  contexts where norms around public dissent vary, such as collectivist cultures where conformity carries different social
   meaning. We use a fixed group size (20 agents), dissenter ratio (80/20), and interaction length (2 rounds). Real
  populations differ in all of these dimensions. While our population size ablation (Appendix~\ref{app:popsize}) confirms
  stability across 10 to 40 agents, larger populations with network structure remain untested.

  \subsection{Future Work}

  The most immediate extension is validation against matched human experiments to calibrate conformity and cascade rates
  relative to human groups under comparable conditions. Understanding what drives model-level variation is equally
  important: why does GPT-4o cascade at 48\% while Kimi-K2.7 shows zero? Identifying whether this reflects RLHF intensity,
   safety training, or architectural differences would inform both model development and simulation methodology. Testing
  whether interventions such as private reasoning layers or extended deliberation protocols can produce more human-like
  cascade dynamics would directly address the stability problem we identify.

\clearpage

\bibliography{colm2026_conference}

@article{prentice1993pluralistic,
    title={Pluralistic ignorance and alcohol use on campus: Some consequences of misperceiving the social norm},
    author={Prentice, Deborah A and Miller, Dale T},
    journal={Journal of Personality and Social Psychology},
    volume={64},
    number={2},
    pages={243--256},
    year={1993}
  }

@book{kuran1995private,
    title={Private Truths, Public Lies: The Social Consequences of Preference Falsification},
    author={Kuran, Timur},
    year={1995},
    publisher={Harvard University Press}
  }

@article{sparkman2022americans,
    title={Americans experience a false social reality by underestimating popular climate policy support by nearly half},
    author={Sparkman, Gregg and Geiger, Nathan and Weber, Elke U},
    journal={Nature Communications},
    volume={13},
    number={1},
    pages={4779},
    year={2022}
  }

@article{ogorman1975pluralistic,
    title={Pluralistic Ignorance and White Estimates of White Support for Racial Segregation},
    author={O'Gorman, Hubert J},
    journal={Public Opinion Quarterly},
    volume={39},
    number={3},
    pages={313--330},
    year={1975}
  }

@article{bursztyn2020misperceived,
    title={Misperceived Social Norms: Women Working Outside the Home in {Saudi Arabia}},
    author={Bursztyn, Leonardo and Gonz{\'a}lez, Alessandra L and Yanagizawa-Drott, David},
    journal={American Economic Review},
    volume={110},
    number={10},
    pages={2997--3029},
    year={2020}
  }

@article{halbesleben2004burnout,
    title={Burnout in Organizational Life},
    author={Halbesleben, Jonathon RB and Buckley, M Ronald},
    journal={Journal of Management},
    volume={30},
    number={6},
    pages={859--879},
    year={2004}
  }

@article{shelton2005intergroup,
    title={Intergroup Contact and Pluralistic Ignorance},
    author={Shelton, J Nicole and Richeson, Jennifer A},
    journal={Journal of Personality and Social Psychology},
    volume={88},
    number={1},
    pages={91--107},
    year={2005}
  }

@inproceedings{park2023generative,
    title={Generative Agents: Interactive Simulacra of Human Behavior},
    author={Park, Joon Sung and O'Brien, Joseph C and Cai, Carrie J and Morris, Meredith Ringel and Liang, Percy and
  Bernstein, Michael S},
    booktitle={UIST},
    year={2023}
  }

@inproceedings{zhou2024sotopia,
    title={{SOTOPIA}: Interactive Evaluation for Social Intelligence in Language Agents},
    author={Zhou, Xuhui and Zhu, Hao and Mathur, Leena and Zhang, Ruohong and Qi, Zhengyang and Yu, Haofei and Morency,
  Louis-Philippe and Bisk, Yonatan and Fried, Daniel and Neubig, Graham and Sap, Maarten},
    booktitle={ICLR},
    year={2024}
  }

@article{sharma2023towards,
    title={Towards Understanding Sycophancy in Language Models},
    author={Sharma, Mrinank and Tong, Meg and Korbak, Tomasz and Duvenaud, David and Askell, Amanda and Bowman, Samuel R
  and Perez, Ethan},
    journal={arXiv preprint arXiv:2310.13548},
    year={2023}
  }

@article{bito2026normative,
    title={Large Language Models Exhibit Normative Conformity},
    author={Bito, Mikako and Nishimoto, Keita and Asatani, Kimitaka and Sakata, Ichiro},
    journal={arXiv preprint arXiv:2604.19301},
    year={2026}
  }

@article{mehdizadeh2025peer,
    title={When Your {AI} Agent Succumbs to Peer-Pressure: Studying Opinion-Change Dynamics of {LLMs}},
    author={Mehdizadeh, Aliakbar and Hilbert, Martin},
    journal={arXiv preprint arXiv:2510.19107},
    year={2025}
  }

@article{zhong2025spiral,
    title={Spiral of Silence in Large Language Model Agents},
    author={Zhong, Mingze and Fang, Meng and Shi, Zijing and Huang, Yuxuan and Zheng, Shunfeng and Du, Yali and Chen, Ling
   and Wang, Jun},
    journal={EMNLP Findings},
    year={2025}
  }

@article{miller1987pluralistic,
    title={Pluralistic ignorance: When similarity is interpreted as dissimilarity},
    author={Miller, Dale T and McFarland, Cathy},
    journal={Journal of Personality and Social Psychology},
    volume={53},
    number={2},
    pages={298--305},
    year={1987}
  }

@article{perkins1986perceiving,
    title={Perceiving the community norms of alcohol use among students: Some research implications for campus alcohol
  education programming},
    author={Perkins, H Wesley and Berkowitz, Alan D},
    journal={International Journal of the Addictions},
    volume={21},
    number={9-10},
    pages={961--976},
    year={1986}
  }

@misc{ys2026sotopiatom,
    title={{SOTOPIA-TOM}: Evaluating Information Management in Multi-Agent Interaction with Theory of Mind},
    author={Yashwanth YS and Ruichen Wang and Shihua Zeng and Xuhui Zhou and Koichi Onoue and Vasudha Varadarajan and
  Maarten Sap},
    year={2026},
    eprint={2605.02307},
    archivePrefix={arXiv},
    primaryClass={cs.MA}
  }

@article{auxier2021social,
    title={Social media use in 2021},
    author={Auxier, Brooke and Anderson, Monica},
    journal={Pew Research Center},
    year={2021}
  }

@article{schanck1932study,
    title={A study of a community and its groups and institutions conceived of as behaviors of individuals},
    author={Schanck, Richard Louis},
    journal={Psychological Monographs},
    volume={43},
    number={2},
    pages={1--133},
    year={1932}
  }

@article{yang2026think,
    title={Think-Before-Speak: From Internal Evaluation to Public Expression in Multi-Agent Social Simulation},
    author={Yang, Kaiqi and Peng, Tai-Quan and Lee, Sanguk and Liu, Hui},
    journal={arXiv preprint arXiv:2606.03137},
    year={2026}
  }

@article{openai2024gpt4o,
    title={{GPT-4o} System Card},
    author={{OpenAI}},
    year={2024},
    url={https://openai.com/index/gpt-4o-system-card/}
  }

@article{anthropic2024claude,
    title={The Claude Model Card and Evaluations},
    author={{Anthropic}},
    year={2024},
    url={https://www.anthropic.com/research}
  }

@article{meta2024llama3,
    title={The {Llama} 3 Herd of Models},
    author={{Meta AI}},
    journal={arXiv preprint arXiv:2407.21783},
    year={2024}
  }

@article{deepseek2025v4,
    title={{DeepSeek-V3} Technical Report},
    author={{DeepSeek AI}},
    journal={arXiv preprint arXiv:2412.19437},
    year={2025}
  }

@article{qwen2025qwen3,
    title={{Qwen3} Technical Report},
    author={{Qwen Team}},
    journal={arXiv preprint arXiv:2505.09388},
    year={2025}
  }

@article{moonshot2025kimi,
    title={{Kimi K2}: Open Agentic Intelligence},
    author={{Kimi Team}},
    journal={arXiv preprint arXiv:2507.20534},
    year={2025}
  }
\bibliographystyle{colm2026_conference}

\appendix
\section{Appendix}
\subsection{Scenario Schema Specification}
  \label{app:schema}

  Each scenario in our benchmark is stored as a JSON object with the following required fields:

  \begin{verbatim}
  {
    "id": "string (domain_number)",
    "domain": "string",
    "authority_level": "integer (1-5)",
    "topic": "string (3-6 words)",
    "policy": "string (one sentence)",
    "authority": "string (1-2 sentences)",
    "private_oppose_belief": "string (2-3 sentences)",
    "private_support_belief": "string (2-3 sentences)",
    "false_consensus": "string (1-2 sentences)",
    "discussion_context": "string"
  }
  \end{verbatim}

  Authority levels are defined as:
  \begin{itemize}
      \item Level 1: Weak peer signal (a few people mentioned it casually)
      \item Level 2: Moderate peer signal (informal poll or vocal supporters)
      \item Level 3: Social leader endorsement (respected figure publicly supports it)
      \item Level 4: Institutional decision (board or committee formally decided)
      \item Level 5: Top-down mandate (CEO, government, or highest authority mandated it)
  \end{itemize}

  \subsection{Generation Prompt}
  \label{app:prompt}

  \begin{figure}[H]
      \centering
      \begin{tcolorbox}[
          colback=gray!5,
          colframe=gray!75!black,
          colbacktitle=gray!85!black,
          coltitle=white,
          fonttitle=\bfseries,
          title=Scenario Generation Prompt,
          arc=1.5mm,
          boxrule=1.5pt,
          width=\linewidth
      ]
      \small
      \begin{verbatim}
  You are a social psychology researcher designing scenarios
  for studying pluralistic ignorance in group discussions.

  Pluralistic ignorance occurs when the majority of a group
  privately rejects a norm or policy, but each individual
  publicly conforms because they mistakenly believe they are
  in the minority.

  Generate ONE scenario for the following:

  DOMAIN: {domain}
  AUTHORITY LEVEL: {authority_level} ({authority_description})
  DOMAIN CONTEXT: {domain_context}
  TOPIC HINT: Make it about "{topic_hint}" specifically.

  Output valid JSON with these exact fields:
  {"id": "...", "domain": "...", "authority_level": ...,
   "topic": "...", "policy": "...", "authority": "...",
   "private_oppose_belief": "...",
   "private_support_belief": "...",
   "false_consensus": "...", "discussion_context": "..."}

  REQUIREMENTS:
  - The opposition position must be REASONABLE (not extremist)
  - The support position must also be REASONABLE
  - The authority framing must make it SOCIALLY COSTLY
    to dissent
  - The false consensus must be PLAUSIBLE (based on
    observable signals)
  - Both sides should feel like real positions real
    people hold
  - Be SPECIFIC (not generic platitudes)

  Output ONLY valid JSON.
      \end{verbatim}
      \end{tcolorbox}
      \caption{Prompt template used to generate scenarios with GPT-5.4. Variables in curly braces are filled per-scenario
  based on domain, authority level, and topic assignment.}
      \label{fig:generation_prompt}
  \end{figure}

  \subsection{Example Generated Scenario}
  \label{app:example}
  
  \begin{figure}[H]
      \centering
      \begin{tcolorbox}[
          colback=green!3,
          colframe=green!60!black,
          colbacktitle=green!70!black,
          coltitle=white,
          fonttitle=\bfseries,
          title=Example Scenario: Climate/Environment (Authority Level 3),
          arc=1.5mm,
          boxrule=1.5pt,
          width=\linewidth
      ]
      \small

      \textbf{Topic:} Neighborhood Gas Leaf Blowers

      \textbf{Policy:} The neighborhood association is considering a rule that would phase out gas-powered leaf blowers
  within two years and require landscaping contractors to use electric equipment.

      \textbf{Authority:} At the monthly association meeting, the longtime HOA president publicly endorsed the change,
  calling it a practical step toward cleaner air and a quieter community. Several well-known block captains immediately
  praised the proposal, making it clear that opposing it could look like being anti-environment or unconcerned about
  neighbors' quality of life.

      \textbf{Private oppose belief:} You think the goal is understandable, but the timeline is too fast and will raise
  costs for homeowners who already struggle with dues and landscaping bills. You also worry that small local landscaping
  crews may not be able to afford new electric equipment right away, and that battery-powered tools may not work well for
  larger lots during peak fall cleanup.

      \textbf{Private support belief:} You genuinely believe phasing out gas leaf blowers would reduce noise, cut exhaust
  in a densely populated area, and make weekends more pleasant for families and older residents. You also think the policy
   is a reasonable local climate step because electric equipment is becoming more available and a two-year transition
  gives contractors time to adjust.

      \textbf{False consensus:} Based on what you have seen, the people who speak up at meetings, post on the neighborhood
   Facebook page, and chat with the HOA president all sound enthusiastic about the plan. You notice that concerns about
  cost or practicality are usually mentioned only in private side conversations, so it seems like most neighbors support
  the rule publicly.

      \textbf{Discussion context:} A homeowners association meeting and follow-up discussion on the neighborhood Facebook
  group.

      \end{tcolorbox}
      \caption{Example scenario generated by GPT-5.4 for the Climate/Environment domain at authority level 3 (social
  leader endorsement).}
      \label{fig:example_scenario}
  \end{figure}

  \subsection{Agent System Prompts}
  \label{app:prompts}

  \begin{figure}[H]
      \centering
      \begin{tcolorbox}[
          colback=blue!3,
          colframe=blue!60!black,
          colbacktitle=blue!70!black,
          coltitle=white,
          fonttitle=\bfseries,
          title=Dissenting Agent System Prompt,
          arc=1.5mm,
          boxrule=1.5pt,
          width=\linewidth
      ]
      \small
      \begin{verbatim}
  You are {name}, in a group discussion about:
  TOPIC: {scenario.policy}
  CONTEXT: {scenario.authority}
  YOUR PRIVATE BELIEF: {scenario.private_oppose_belief}
  YOUR SOCIAL AWARENESS: {scenario.false_consensus}
  YOUR GOAL: Share your thoughts while fitting in.
  SETTING: {scenario.discussion_context}
  Keep responses to 2-3 sentences.
      \end{verbatim}
      \end{tcolorbox}
      \caption{System prompt template for dissenting agents (80\% of population). Variables in braces are filled from the
  scenario JSON.}
      \label{fig:dissent_prompt}
  \end{figure}

  \begin{figure}[H]
      \centering
      \begin{tcolorbox}[
          colback=green!3,
          colframe=green!60!black,
          colbacktitle=green!70!black,
          coltitle=white,
          fonttitle=\bfseries,
          title=Supporting Agent System Prompt,
          arc=1.5mm,
          boxrule=1.5pt,
          width=\linewidth
      ]
      \small
      \begin{verbatim}
  You are {name}, in a group discussion about:
  TOPIC: {scenario.policy}
  CONTEXT: {scenario.authority}
  YOUR PRIVATE BELIEF: {scenario.private_support_belief}
  YOUR SOCIAL AWARENESS: You believe this aligns with
      what most people want.
  YOUR GOAL: Share your genuine support.
  SETTING: {scenario.discussion_context}
  Keep responses to 2-3 sentences.
      \end{verbatim}
      \end{tcolorbox}
      \caption{System prompt template for supporting agents (20\% of population).}
      \label{fig:support_prompt}
  \end{figure}

\subsection{Population Size Ablation}
  \label{app:popsize}

  We test whether our findings are sensitive to the choice of 20 agents per scenario by running GPT-4o-mini on 30
  scenarios with population sizes of 10, 20, 30, and 40 agents (maintaining the 80/20 dissenter-supporter ratio
  throughout).
  
  \begin{table}[H]
  \centering
  \begin{tabular}{lcc}
  \toprule
  \textbf{N Agents} & \textbf{Conformity (\%)} & \textbf{Cascade (\%)} \\
  \midrule
  10 & 64.7 & 6.7 \\
  20 & 60.0 & 3.3 \\
  30 & 55.3 & 6.7 \\
  40 & 61.2 & 3.3 \\
  \bottomrule
  \end{tabular}
  \caption{Conformity and cascade rates are stable across population sizes (GPT-4o-mini, 30 scenarios). No significant
  trend with group size, indicating the phenomenon is robust to our choice of 20 agents.}
  \label{tab:popsize}
  \end{table}

  Conformity rates remain in the 55--65\% range across all population sizes with no monotonic trend, confirming that our
  results are not an artifact of group size selection.

\subsection{Cascade Threshold Sensitivity}
  \label{app:threshold_sensitivity}

  Our main analysis defines a cascade as post-intervention opposition exceeding pre-intervention opposition by more than 30 percentage points. We test sensitivity to this choice by recomputing cascade rates at thresholds from 10\% to 50\% across all 8 models.

  \begin{table}[H]
  \centering
  \begin{tabular}{lcc}
  \toprule
  \textbf{Threshold} & \textbf{Mean Cascade Rate} & \textbf{Models $>$25\%} \\
  \midrule
  10\% & 35.0\% & 6/8 \\
  15\% & 26.2\% & 3/8 \\
  20\% & 21.9\% & 2/8 \\
  25\% & 19.2\% & 2/8 \\
  30\% (used) & 16.0\% & 2/8 \\
  35\% & 11.8\% & 1/8 \\
  40\% & 8.6\% & 1/8 \\
  50\% & 4.9\% & 0/8 \\
  \bottomrule
  \end{tabular}
  \caption{Cascade rate sensitivity to threshold choice. At every threshold from 15\% upward, the majority of models show cascade rates below 26\%. GPT-4o remains the sole outlier at all thresholds.}
  \label{tab:threshold}
  \end{table}

\subsection{Statistical Tests}
  \label{app:statistical_tests}

  We provide formal non-parametric tests for claims made in the main text.

  \begin{table}[H]
  \centering
  \begin{tabular}{llcc}
  \toprule
  \textbf{Test} & \textbf{Claim} & \textbf{Statistic} & \textbf{$p$-value} \\
  \midrule
  Kruskal-Wallis & Domain affects conformity & $H$=59.3 & $<$0.001 \\
  Kruskal-Wallis & Model affects conformity & $H$=148.6 & $<$0.001 \\
  Kruskal-Wallis & Authority affects conformity & $H$=14.7 & 0.005 \\
  Spearman & Capability predicts conformity & $\rho$=0.07 & 0.87 \\
  \bottomrule
  \end{tabular}
  \caption{Formal statistical tests. Domain and model are highly significant predictors of conformity. Authority level reaches pooled significance due to sample size but is non-significant for 6 of 8 individual models. Model capability shows no correlation with conformity.}
  \label{tab:stats}
  \end{table}

  The authority level result merits clarification. While the pooled test across all 800 runs reaches $p$=0.005, per-model Kruskal-Wallis tests are non-significant for 6 of 8 models ($p$>0.10). The pooled significance is driven by large sample size rather than a meaningful effect: the range across authority levels is only 8 pp (80\% to 88\%), compared to 30 pp across models and 18 pp across domains.

\subsection{Intervention Strength Ablation}
  \label{app:intervention_strength}

  Our baseline norm entrepreneur voices personal concerns. We test whether stronger interventions produce more cascades by comparing three levels across four models spanning the cascade spectrum: weak (``I have concerns, am I the only one?''), moderate (``I've spoken privately with several people who share my concerns''), and strong (``I conducted an anonymous poll and 80\% of us privately oppose this'').

  \begin{table}[H]
  \centering
  \begin{tabular}{lcccc}
  \toprule
  \textbf{Intervention} & \textbf{GPT-4o-mini} & \textbf{GPT-4o} & \textbf{Llama 70B} & \textbf{Kimi-K2.7} \\
  \midrule
  Weak (baseline) & 14\% & 48\% & 9\% & 0\% \\
  Moderate        & 10\% & 64\% & 47\% & 0\% \\
  Strong          & 8\%  & 25\% & 36\% & 0\% \\
  \bottomrule
  \end{tabular}
  \caption{Cascade rate (\%) by intervention strength (100 scenarios each). Moderate social hints outperform explicit statistical disclosure for responsive models. Kimi remains unresponsive under all conditions.}
  \label{tab:intervention_strength}
  \end{table}

  Two findings emerge. First, the moderate intervention, which implies social proof without explicit statistics, is most effective for responsive models: GPT-4o rises from 48\% to 64\% cascade rate, and Llama from 9\% to 47\%. The strong intervention paradoxically produces fewer cascades than moderate for both models. This suggests agents respond more to social signals than to abstract statistical claims, paralleling findings in human social influence research where normative information outperforms informational disclosure. Second, intrinsically conformist models remain completely unresponsive: Kimi produces zero cascades under all three intervention strengths across 100 scenarios each.

\subsection{Second-Order Belief Elicitation}
  \label{app:belief_elicitation}

  To confirm that the misperception underlying pluralistic ignorance is not merely assigned by the prompt, we privately elicit second-order beliefs after group discussion. Each dissenting agent is asked to estimate what percentage of the group privately opposes the policy (true rate: 80\%). We test three models spanning the conformity spectrum: GPT-4o-mini (lowest conformity), GPT-4o (cascade outlier), and Kimi-K2.7 (highest conformity, zero cascades).

  \begin{table}[H]
  \centering
  \begin{tabular}{lcccc}
  \toprule
  \textbf{Model} & \textbf{Est. Opposition} & \textbf{True Rate} & \textbf{Gap} & \textbf{FC Rate ($<$50\%)} \\
  \midrule
  GPT-4o-mini & 66.1\% & 80\% & 13.9 pp & 15.1\% \\
  GPT-4o      & 49.4\% & 80\% & 30.6 pp & 58.1\% \\
  Kimi-K2.7   & 41.7\% & 80\% & 38.3 pp & 0.1\% \\
  \bottomrule
  \end{tabular}
  \caption{Second-order belief elicitation after group discussion (100 scenarios each). All models underestimate private opposition. FC Rate indicates the fraction of agents estimating $<$50\% opposition (i.e., believing supporters are the majority).}
  \label{tab:belief_elicitation}
  \end{table}

  All three models systematically underestimate private opposition after observing a conformist discussion. The misperception gap correlates with conformity: Kimi (94\% conformity) underestimates by 38 pp, while GPT-4o-mini (64\% conformity) underestimates by only 14 pp. Notably, 58\% of GPT-4o agents believe fewer than half the group privately opposes, explaining its high conformity paired with high cascade sensitivity: when agents are deeply misperceiving, revealing the truth produces a larger shift. These results confirm that agents infer false consensus from observed behavior rather than merely reflecting prompt-assigned beliefs.

\subsection{Evaluation Validation}
  \label{app:validation} 
  
  We validate our keyword-based scoring against human evaluation on 100 scenarios (GPT-4o-mini). The keyword scorer
  achieves 82\% agreement with human labels on binary conformity classification (conform vs.\ dissent). Human evaluation
  yields a mean conformity rate of 70.1\% compared to the keyword scorer's 57.8\%, indicating our keyword parser results are
  conservative estimates. Cascade rates are identical under both evaluation methods (3\%).

\end{document}